# Analytical design of flat-top transmission filters composed of several resonant structures

Leonid L. Doskolovich[1,2,*], Nikita V. Golovastikov[1,2], Dmitry A. Bykov[1,2], and Evgeni A. Bezus[1,2]

[1] Image Processing Systems Institute — Branch of the Federal Scientific Research Centre "Crystallography and Photonics" of Russian Academy of Sciences, 151 Molodogvardeyskaya st., Samara 443001, Russia
[2] Samara National Research University, 34 Moskovskoe shosse, Samara 443086, Russia
*leonid@smr.ru

**Abstract:** Resonant properties of composite structures consisting of several identical resonant diffractive structures (e.g. multilayer thin-film structures or guided-mode resonance gratings) separated by phase-shift layers are investigated theoretically. Using the scattering matrix formalism, we analytically demonstrate that, at properly chosen thicknesses of the phase-shift layers, the composite structures comprising two or four resonant diffractive structures with a Lorentzian transmittance profile optically implement the Butterworth filters of the order two or three, respectively, and enable achieving flat-top transmission spectra with steep slopes and low sidebands. In addition, we show that the composite structures consisting of three or four second-order Butterworth filters can accurately approximate the fourth- or fifth-order Butterworth filters, respectively. The presented theoretical results are confirmed by rigorous numerical simulations of composite structures consisting of the so-called W-structures (simple three-layer resonant structures comprising a high-index core layer and two low-index cladding layers in a high-index dielectric environment). The simulation results confirm the formation of flat-top transmittance peaks, the shape of which fully agrees with the derived theoretical description. Moreover, we demonstrate an exceptionally simple mechanism of controlling the transmittance peak width, which consists in changing the thicknesses of the cladding layers of the initial W-structure and enables generating flat-top transmission peaks with a significantly subnanometer size.

## 1. Introduction

Optical filters that selectively reflect or transmit input light are ubiquitously used in spectroscopic measurements, laser systems, optical communications, displays, hyperspectral imaging and many other practical applications. As optical filters, resonant diffractive structures are widely used, including various thin-film structures [1], subwavelength-period guided-mode resonance gratings (GMRGs) [2–7], and "combined" multilayer structures containing GMRGs embedded into multilayer thin-films [7–10].

The simplest thin-film reflecting filter is the Bragg grating (BG), which corresponds to a periodic quarter-wave layer system. A phase-shifted Bragg grating (PSBG) consists of two symmetric BGs separated by a half-wave spacer ("defect" or "phase-shift" layer) and enables obtaining a narrow transmission passband centered at the design wavelength [1]. In the early works dedicated to the GMRG-based filters, only the filters operating in reflection were proposed [2–4, 8]. The utilization of combined multilayer structures (i.e. GMRGs in combination with multilayer thin-film structures) made it possible to simplify the filter configuration and decrease the required number of layers. The first transmission bandpass filters corresponding to such combined multilayer structures were presented in 1995 [9]. More advanced GMRG-based transmission filters include only a few thin-film layers or even a single-layer GMRG [6, 7, 10].

As a rule, transmission filters based on PSBGs or GMRGs enable obtaining a Lorentzian transmittance peak. At the same time, the Lorentzian line shape does not always fit the design requirements. For many applications, achieving a nearly rectangular transmittance spectrum with flat top and steep slopes is highly desirable [1, 11, 12].

Currently, flat-top narrow-band filters operating in transmission are mostly based on thin-film Bragg structures containing multiple defects (i.e. phase-shift spacers breaking the grating periodicity) [1]. In such structures, several leaky modes localized at the phase-shift regions may exist. The existence of several modes makes the line-shape of the resonance significantly more complex and makes it possible to achieve a flat-top transmittance peak by optimizing the parameters of the phase-shift spacers [1, 11]. A similar approach can be applied to the design of GMRG-based transmission filters. In particular, Ref. [12] considers the design of flat-top bandpass filters consisting of several GMRGs separated by phase-shift spacers. Let us note that, to the best of our knowledge, no theoretical description of the formation of a flat-top transmittance peak has been presented in the published works. The parameters of the multilayer thin-film structures and GMRG-based structures are, as a rule, determined using numerical optimization.

In this work, we present a theoretical analysis of the spectra of composite structures consisting of several identical resonant diffractive structures (RDSs) with a Lorentzian transmittance profile. Using the scattering matrix formalism, we analytically demonstrate that composite structures comprising two or four RDSs optically implement the Butterworth filters of the order two or three, respectively, and enable achieving flat-top transmittance spectra with steep slopes and no sidelobes. In addition, we show that composite structures consisting of three or four second-order Butterworth filters approximate the Butterworth filters of the order four or five, respectively.

The obtained theoretical results are confirmed by rigorous numerical simulations. As the resonant building block of the composite structure, the so-called W-structure (or W-type waveguide) [13–16] is utilized. The W-structure is one of the simplest resonant structures supporting leaky modes, which consists of only three dielectric layers in a dielectric surrounding medium. The presented simulation results of the designed composite structures confirm the formation of flat-top transmittance peaks, the shape of which fully agrees with the derived theoretical description. Moreover, an exceptionally simple mechanism of controlling the transmittance peak width is demonstrated, which consists in changing the thicknesses of the cladding layers of the initial W-structure.

**2. Theoretical analysis of the spectra of composite structures**

In this Section, we consider composite structures consisting of several identical resonant dielectric structures. As such structures, guided-mode resonance gratings, phase-shifted Bragg gratings or other resonance thin-film structures can be used. For example, Fig. 1(c) shows a composite structure comprising two so-called W-structures (or W-type waveguides) [13–16]. As mentioned above, the W-structure is the simplest resonant structure supporting leaky modes. W-structures consist of three dielectric layers: a high-index core layer and two low-index cladding layers [Fig. 1(a)]. The structure is supposed to be located in a high-index symmetric dielectric environment. The term "W-structure" is due to the fact that the refractive index profile of the structure resembles the letter "W" [see Fig. 1(b)].

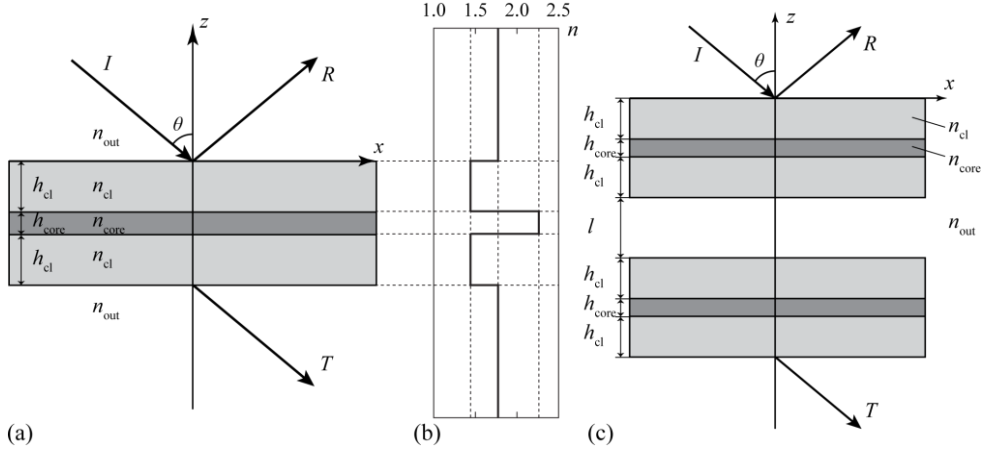

Fig. 1. Three-layer W-structure (a), refractive index profile of the W-structure (b), and a composite structure consisting of two W-structures (c).

Let us suppose that the lossless RDSs constituting the composite structure have a Lorentzian transmittance spectrum. In this case, the complex transmission and reflection coefficients of the RDS considered as functions of the angular frequency $\omega$ of the incident light can be approximated by the following expressions:

$$r_1(\omega) = \exp\{i\varphi\}\frac{\omega - \operatorname{Re}\omega_p}{\omega - \omega_p}, \quad t_1(\omega) = \exp\{i\varphi\}\frac{i\operatorname{Im}\omega_p}{\omega - \omega_p}, \qquad (1)$$

where the pole $\omega_p$ is the complex frequency of the eigenmode of the resonant structure. The form of the coefficients $r_1(\omega)$ and $t_1(\omega)$ given by Eq. (1) enforces the energy conservation law $|r_1(\omega)|^2 + |t_1(\omega)|^2 = 1$. According to Eq. (1), at $\omega_0 = \operatorname{Re}\omega_p$ the reflectance $R_1(\omega) = |r_1(\omega)|^2$ vanishes and the transmittance $T_1(\omega) = |t_1(\omega)|^2$ reaches unity. The width of the resonance is determined by the imaginary part of the pole $\omega_p$ and can be described by the quality factor $Q = \operatorname{Re}\omega_p / 2|\operatorname{Im}\omega_p|$. Indeed, from Eq. (1), it is easy to obtain that the half-width of the resonant transmittance peak $T_1(\omega)$ (reflectance dip $R_1(\omega)$) at the 0.5 level amounts to $\Delta_1 = |\operatorname{Im}\omega_p|$.

For the analysis of composite structures, it is convenient to describe the optical properties of an RDS at a fixed incidence angle $\theta$ by the scattering matrix [17–20]

$$\mathbf{S}_1(\omega) = \begin{pmatrix} t_1(\omega) & r_1(\omega) \\ r_1(\omega) & t_1(\omega) \end{pmatrix}. \qquad (2)$$

The matrix $\mathbf{S}_1(\omega)$ relates the complex amplitudes of the plane waves $i_u(\omega)$ and $i_d(\omega)$, which impinge on the RDS from the superstrate and the substrate regions, respectively, with the complex amplitudes of the reflected and transmitted waves $r(\omega)$ and $t(\omega)$:

$$\begin{bmatrix} t(\omega) \\ r(\omega) \end{bmatrix} = \mathbf{S}_1(\omega) \begin{bmatrix} i_u(\omega) \\ i_d(\omega) \end{bmatrix}. \qquad (3)$$

Let us note that the symmetric form of the scattering matrix $\mathbf{S}_1(\omega)$ given by Eq. (2) assumes that the RDS possesses a horizontal symmetry plane. The existence of such a symmetry plane is necessary for the reflection coefficient $r_1(\omega)$ given by Eq. (1) to vanish at a real frequency [20].

Let us consider a composite structure consisting of two identical RDSs described by the scattering matrix of Eq. (2) and separated by a phase-shift layer with the thickness $l_1$. In this case, the scattering matrix of the composite structure can be expressed through the matrix $\mathbf{S}_1(\omega)$ in the form [17, 21]

$$\mathbf{S}_2(\omega) = \mathbf{S}_1(\omega) * \mathbf{L}(l_1) * \mathbf{S}_1(\omega), \tag{4}$$

where $*$ denotes the Redheffer star product:

$$\begin{pmatrix} a_{1,1} & a_{1,2} \\ a_{2,1} & a_{2,2} \end{pmatrix} * \begin{pmatrix} b_{1,1} & b_{1,2} \\ b_{2,1} & b_{2,2} \end{pmatrix} =$$
$$= \frac{1}{1 - a_{1,2} b_{2,1}} \begin{pmatrix} b_{1,1} a_{1,1} & b_{1,2} - a_{1,2}(b_{1,2} b_{2,1} - b_{1,1} b_{2,2}) \\ a_{2,1} - b_{2,1}(a_{1,2} a_{2,1} - a_{1,1} a_{2,2}) & a_{2,2} b_{2,2} \end{pmatrix}, \tag{5}$$

and $\mathbf{L}(l_1)$ is the scattering matrix of the phase-shift layer. The scattering matrix $\mathbf{L}(l_1)$ has the form

$$\mathbf{L}(l_1) = \exp\{i\psi(l_1)\} \mathbf{E}, \tag{6}$$

where $\psi(l_1)$ is the phase shift acquired by the plane wave upon propagation through a homogeneous dielectric layer with thickness $l_1$ and refractive index $n$, and $\mathbf{E}$ is the 2x2 identity matrix. For simplicity, here we assume that $\psi(l_1)$ is frequency-independent and can be approximated as

$$\psi(l_1) = l_1 \sqrt{(n\omega_0/c)^2 - k_x^2}, \tag{7}$$

where $\omega_0 = \operatorname{Re}\omega_p$ is the resonant frequency, $k_x = (\omega_0/c) n_{out} \sin\theta$ is the tangential component of the wave vectors of the plane waves impinging on the structure from the superstrate and substrate regions, $c$ is the free-space speed of light, $\theta$ is the angle of incidence, and $n_{out}$ is the refractive index of both the superstrate and substrate regions.

By using the definition of the Redheffer product given by Eq. (5) in Eq. (4), one can obtain the scattering matrix of the composite structure in the form

$$\mathbf{S}_2(\omega) = \begin{pmatrix} t_2(\omega) & r_2(\omega) \\ r_2(\omega) & t_2(\omega) \end{pmatrix} =$$
$$= \frac{1}{1 - \exp\{2i\psi\} r_1^2} \begin{pmatrix} \exp\{2i\psi\} t_1^2 & r_1\left[1 - \exp\{2i\psi\}(r_1^2 - t_1^2)\right] \\ r_1\left[1 - \exp\{2i\psi\}(r_1^2 - t_1^2)\right] & \exp\{2i\psi\} t_1^2 \end{pmatrix}. \tag{8}$$

Substituting Eq. (1) into Eq. (8), we obtain the following resonant approximations of the reflection and transmission coefficients of the composite structure:

$$r_2(\omega) = \exp\{i\varphi\} \frac{(\omega - \operatorname{Re}\omega_p)\left[\omega_p - \omega + \sigma(l_1)(\omega - \omega_p^*)\right]}{-(\omega - \omega_p)^2 + \sigma(l_1)(\omega - \operatorname{Re}\omega_p)^2},$$

$$t_2(\omega) = \frac{\sigma(l_1) \exp\{i\varphi\} (\operatorname{Im}\omega_p)^2}{-(\omega - \omega_p)^2 + \sigma(l_1)(\omega - \operatorname{Re}\omega_p)^2}, \tag{9}$$

where $\sigma(l_1) = \exp\{2i[\varphi + \psi(l_1)]\}$. The expressions for the reflection coefficient $r_2(\omega)$ and the transmission coefficient $t_2(\omega)$ given by Eq. (9) have a significantly more complex form than the corresponding coefficients $r_1(\omega)$ and $t_1(\omega)$ of the initial RDS. In particular, the composite structure consisting of two RDSs supports two eigenmodes. The complex frequencies of these modes are the roots of the denominator of both expressions in Eq. (9), which is a second-order polynomial of the frequency $\omega$. In addition, the reflection coefficient $r_2(\omega)$ has two zeros corresponding to the roots of the polynomial being the numerator of the corresponding expression in Eq. (9).

*2.1. Butterworth filter of the second order*

A more complex form of the coefficients $r_2(\omega)$ and $t_2(\omega)$ gives an opportunity to control the shapes of the reflectance and transmittance spectra. In particular, if the thickness of the phase-shift layer $l_1$ satisfies the condition

$$\psi(l_1) = \frac{\pi}{2} - \varphi + \pi m, \; m \in \mathbb{N}, \tag{10}$$

where the phase-shift $\psi(l_1)$ is defined by Eq. (7) and $\varphi$ is the phase of the transmission coefficient $t_1(\omega_0)$ [Eq. (1)], the coefficients $r_2(\omega)$ and $t_2(\omega)$ take the form

$$r_2(\omega) = \frac{\exp\{i\varphi\}(\omega - \operatorname{Re}\omega_p)^2}{(\omega - \omega_{p,1})(\omega - \omega_{p,2})}, \quad t_2(\omega) = \frac{-i\exp\{i\varphi\}(\operatorname{Im}\omega_p)^2}{2(\omega - \omega_{p,1})(\omega - \omega_{p,2})}, \tag{11}$$

where

$$\omega_{p,1} = \operatorname{Re}\omega_p + \frac{i-1}{2}\operatorname{Im}\omega_p, \quad \omega_{p,2} = \operatorname{Re}\omega_p + \frac{i+1}{2}\operatorname{Im}\omega_p. \tag{12}$$

Let us note that the reflection coefficient $r_2(\omega)$ has a second-order real-valued zero, whereas the poles of the reflection and transmission coefficients defined by Eq. (12) lie on a circle in the complex plane with the center at the point $\omega_0 = \operatorname{Re}\omega_p$ and the radius $\Delta_2 = \operatorname{Im}\omega_p/\sqrt{2}$. In this case, the expression for $t_2(\omega)$ coincides with the transfer function of the second-order Butterworth filter [22] and, in comparison with $r_1(\omega)$, provides a significantly more rectangular shape of the transmittance peak. Indeed, the squared modulus of the transfer function of the Butterworth filter of the order $M$ reads

$$T_{BW,M}(\omega) = \frac{1}{1 + [(\omega - \omega_0)/\Delta]^{2M}}, \tag{13}$$

where $\omega_0$ and $\Delta$ are the center and radius of the circle, on which the poles lie, respectively. From Eqs. (11) and (12) one can obtain that, according to Eq. (13), the transmittance of the composite structure $T_2(\omega) = |t_2(\omega)|^2$ has the form

$$T_2(\omega) = \frac{1}{1 + [(\omega - \operatorname{Re}\omega_p)/\Delta_2]^4}, \tag{14}$$

where $\Delta_2 = \operatorname{Im}\omega_p/\sqrt{2}$ is the half-width of the resonant transmittance peak of the composite structure at the 0.5 level. Let us note that the half-width $\Delta_2$ is $\sqrt{2}$ times less than that of the initial RDS having a Lorentzian spectral shape.

*2.2. Butterworth filter of the third order*

Next, let us consider a composite structure consisting of four RDSs and corresponding to the superposition of two second-order Butterworth filters. The scattering matrix of such structure has the form

$$\mathbf{S}_4(\omega) = \mathbf{S}_2(\omega) * \mathbf{L}(l_2) * \mathbf{S}_2(\omega), \tag{15}$$

where $l_2$ is the distance (spacer layer thickness) between the two composite structures described by the scattering matrix $\mathbf{S}_2(\omega)$. Note that the coefficients $r_2(\omega)$ and $t_2(\omega)$, which constitute the matrix $\mathbf{S}_2(\omega)$ in Eq. (15), have the form given by Eq. (11). Calculating $\mathbf{S}_4(\omega)$ under the condition

$$\psi(l_2) = \pi - \varphi + \pi m, \, m \in \mathbb{N}, \tag{16}$$

we obtain the reflection and transmission coefficients of the composite structure in the following form:

$$r_4(\omega) = \frac{\exp\{i\varphi\}(\omega - \mathrm{Re}\,\omega_p)^3}{(\omega - \omega_{p,1})(\omega - \omega_{p,2})(\omega - \omega_{p,3})}, \quad t_4(\omega) = \frac{\exp\{i(\pi/2 + \varphi)\}(\mathrm{Im}\,\omega_p)^3}{8(\omega - \omega_{p,1})(\omega - \omega_{p,2})(\omega - \omega_{p,3})}, \tag{17}$$

where the poles $\omega_{p,i}$, $i = 1, 2, 3$ (complex frequencies of the eigenmodes of the composite structures) are

$$\omega_{p,1} = \mathrm{Re}\,\omega_p + \frac{i + \sqrt{3}}{4}\mathrm{Im}\,\omega_p, \quad \omega_{p,2} = \mathrm{Re}\,\omega_p + \frac{i - \sqrt{3}}{4}\mathrm{Im}\,\omega_p,$$
$$\omega_{p,3} = \mathrm{Re}\,\omega_p + \frac{i}{2}\mathrm{Im}\,\omega_p. \tag{18}$$

For this composite structure, the reflection coefficient $r_4(\omega)$ has a real-valued zero of the third order, whereas the poles given by Eq. (18) lie on a circle in the complex plane with the center at the point $\omega_0 = \mathrm{Re}\,\omega_p$ and the radius $\Delta_4 = \mathrm{Im}\,\omega_p/2$. In this case, $t_4(\omega)$ coincides with the transfer function of the third-order Butterworth filter, and the transmittance of the composite structure reads

$$T_4(\omega) = \frac{1}{1 + \left[(\omega - \mathrm{Re}\,\omega_p)/\Delta_4\right]^6}, \tag{19}$$

where the half-width of the transmittance peak $\Delta_4 = \mathrm{Im}\,\omega_p/2$ is already two times less than that of the initial RDS.

*2.3. Approximations of the fourth- and fifth-order Butterworth filters*

Unfortunately, at $N > 2$, the composite structures comprising $N$ second-order Butterworth filters (or $2N$ RDSs) do not exactly implement higher-order Butterworth filters. However, the transmission coefficient of the composite structure consisting of three second-order Butterworth filters is a good approximation of the Butterworth filter of the fourth order. Indeed, the scattering matrix of this structure has the form

$$\mathbf{S}_6(\omega) = \mathbf{S}_2(\omega) * \mathbf{L}(l_2) * \mathbf{S}_2(\omega) * \mathbf{L}(l_2) * \mathbf{S}_2(\omega) = \mathbf{S}_4(\omega) * \mathbf{L}(l_2) * \mathbf{S}_2(\omega). \tag{20}$$

Calculating $\mathbf{S}_6(\omega)$ under the condition of Eq. (16), we obtain the transmission coefficient as

$$t_6(\omega) = \frac{i \exp\{i\varphi\}(\mathrm{Im}\,\omega_p)^4}{32\prod_{i=1}^{4}(\omega - \omega_{p,i})}, \tag{21}$$

where the complex eigenfrequencies of the composite structure $\omega_{p,i}$, $i = 1, 2, 3, 4$ have the form

$$\omega_{p,1} = \mathrm{Re}\,\omega_p + \frac{1}{8}\left(1 + 2i - (1+i)\sqrt{-2 - 5i/2}\right)\mathrm{Im}\,\omega_p,$$
$$\omega_{p,2} = \mathrm{Re}\,\omega_p + \frac{1}{8}\left(1 + 2i + (1+i)\sqrt{-2 - 5i/2}\right)\mathrm{Im}\,\omega_p,$$
$$\omega_{p,3} = \mathrm{Re}\,\omega_p - \frac{1}{8}\left(1 - 2i + (1-i)\sqrt{-2 + 5i/2}\right)\mathrm{Im}\,\omega_p, \tag{22}$$
$$\omega_{p,4} = \mathrm{Re}\,\omega_p - \frac{1}{8}\left(1 - 2i - (1-i)\sqrt{-2 + 5i/2}\right)\mathrm{Im}\,\omega_p.$$

The poles given by Eq. (22) lie in the vicinity of the circle with the center at the point $\omega_0 = \mathrm{Re}\,\omega_p$ and the radius $\Delta_6 = 0.42\,\mathrm{Im}\,\omega_p$. The root-mean-square deviation of the positions of the poles from this circle normalized by the circle radius is only 7.2%. This enables approximating the transmittance $T_6(\omega) = |t_6(\omega)|^2$ by the squared modulus of the fourth-order Butterworth filter:

$$T_6(\omega) \approx \frac{1}{1 + \left[(\omega - \operatorname{Re}\omega_p)/\Delta_6\right]^8}, \qquad (23)$$

where the half-width of the resonant transmittance peak $\Delta_6 = 0.42 \operatorname{Im}\omega_p$ is more than two times less than that of the initial RDS.

Similarly, one can consider a composite structure consisting of four second-order Butterworth filters:

$$\mathbf{S}_8(\omega) = \mathbf{S}_6(\omega) * \mathbf{L}(l_2) * \mathbf{S}_2(\omega). \qquad (24)$$

The transmission coefficient of this structure reads

$$t_8(\omega) = -\frac{\exp\{i\varphi\}(\operatorname{Im}\omega_p)^4}{128 \prod_{i=1}^{5}(\omega - \omega_{p,i})}, \qquad (25)$$

where the poles $\omega_{p,i}$ have the following values:

$$\begin{aligned}
\omega_{p,1} &= \operatorname{Re}\omega_p - (0.43036 + 0.08805\mathrm{i})\operatorname{Im}\omega_p, \\
\omega_{p,2} &= \operatorname{Re}\omega_p + (0.43036 - 0.08805\mathrm{i})\operatorname{Im}\omega_p, \\
\omega_{p,3} &= \operatorname{Re}\omega_p + 0.25(1-\mathrm{i})\operatorname{Im}\omega_p, \\
\omega_{p,4} &= \operatorname{Re}\omega_p - 0.25(1+\mathrm{i})\operatorname{Im}\omega_p, \\
\omega_{p,5} &= \operatorname{Re}\omega_p + 0.32390\mathrm{i}\operatorname{Im}\omega_p.
\end{aligned} \qquad (26)$$

The poles defined by Eq. (26) again lie in the vicinity of the circle with the center at the point $\omega_0 = \operatorname{Re}\omega_p$ and, in this case, the radius $\Delta_8 = 0.38\operatorname{Im}\omega_p$. The normalized root-mean-square deviation of the positions of the poles from this circle is greater than in the previous case and amounts to 14.1%. However, the simulation results presented in the next section demonstrate that the transmittance $T_8(\omega) = |t_8(\omega)|^2$ can still be reasonably well approximated by Eq. (13) at $M = 5$.

## 3. Numerical investigation of composite W-structures

Let us illustrate the theoretical results obtained in the previous section by the results of rigorous numerical simulations of the considered composite structures. As the initial RDS, let us use the W-structure described above [Fig. 1(a)]. Let us remind that this structure consists of three layers: a core layer with the refractive index $n_{\text{core}}$ and thickness $h_{\text{core}}$ and two cladding layers with the refractive index $n_{\text{cl}} < n_{\text{core}}$ and thickness $h_{\text{cl}}$. The structure is located in a symmetric dielectric environment with the refractive index $n_{\text{out}}$. If $n_{\text{out}} > n_{\text{cl}}$, the W-structure supports leaky eigenmodes described by the dispersion relation given in Ref. [13]. For a leaky mode, the eigenfrequency $\omega_p$ is complex. For the considered structure, the quantity $\Delta_1 = |\operatorname{Im}\omega_p|$, which determines the width of the resonance, is an exponentially decaying function of the ratio $\gamma = h_{\text{cl}}/h_{\text{core}}$ [13]. At a large cladding thickness $h_{\text{cl}}$ (at $\gamma \gg 1$), the modes of the W-structure can be approximately described by the dispersion relation of a slab waveguide with the same core layer as in the W-structure and semi-infinite claddings with the refractive index $n_{\text{cl}}$.

As an example, Fig. 2 shows the reflectance and transmittance spectra of the W-structure. The spectra were calculated using a numerically stable implementation of the T-matrix method [23] for an obliquely incident TE-polarized plane wave. The following values of the material and geometrical parameters were used: $n_{\text{core}} = 2.2698\,(\text{TiO}_2)$, $n_{\text{cl}} = 1.457\,(\text{SiO}_2)$, $n_{\text{sup}} = 1.7786\,(\text{SF11})$, $h_{\text{core}} = 40\,\text{nm}$. The simulations were performed for three different thicknesses of the cladding layers: $h_{cl,1} = 700\,\text{nm}$, $h_{cl,2} = 800\,\text{nm}$, and $h_{cl,3} = 1000\,\text{nm}$. The obtained spectra have a Lorentzian line shape and are well described by Eq. (1). Let us mention that the spectra in Fig. 2 corresponding to different cladding thicknesses were calculated at different incidence angles: $\theta_1 = 60.707°$ ($h_{cl,1} = 700\,\text{nm}$), $\theta_2 = 60.713°$ ($h_{cl,2} = 800\,\text{nm}$), and $\theta_3 = 60.715°$ ($h_{cl,3} = 1000\,\text{nm}$). At the angles $\theta_i$, $i = 1,2,3$, the reflection coefficients of the W-structures with the cladding layer thicknesses $h_{cl,i}$, $i = 1,2,3$ vanish at the same frequency $\omega = \omega_0 = 2.811 \cdot 10^{15}\,\text{s}^{-1}$ (free-space wavelength $\lambda_0 = 670\,\text{nm}$).

As noted above, an increase in the cladding layer thickness $h_{cl}$ at a fixed value of $h_{core}$ leads to a decrease in the resonance width $\Delta_1 = |\text{Im}\,\omega_p|$. This effect is clearly visible in Fig. 2. In particular, at the thickness $h_{cl} = h_{cl,3} = 1000\,\text{nm}$, the resonance width is almost 20 times less than at $h_{cl} = h_{cl,1} = 700\,\text{nm}$.

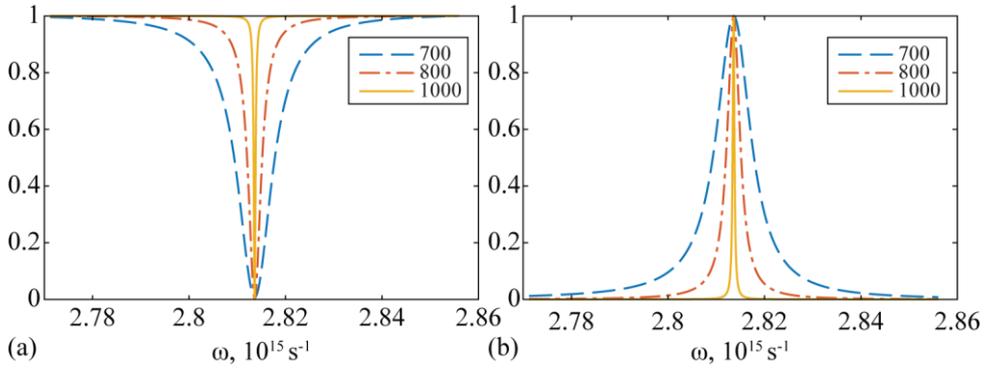

Fig. 2. Reflectance (a) and transmittance (b) spectra of the W-structure at different thicknesses of the cladding layers: $h_{cl,1} = 700\,\text{nm}$ (dashed blue line), $h_{cl,2} = 800\,\text{nm}$ (dash-dotted red line), and $h_{cl,3} = 1000\,\text{nm}$ (solid yellow line).

As the building block for constructing the composite structures, we will use the W-structure with the cladding thickness $h_{cl} = h_{cl,1} = 700\,\text{nm}$. For this structure, we rigorously calculated the complex frequency of the leaky eigenmode $\omega_p = 2.811 \cdot 10^{15} - 4.2 \cdot 10^{12}\,\text{i}\,\text{s}^{-1}$ using the method presented in Ref. [24]. According to Eq. (1), the reflection coefficient vanishes at the frequency $\omega = \omega_0 = \text{Re}\,\omega_p = 2.811 \cdot 10^{15}\,\text{s}^{-1}$. The half-width of the resonant transmittance peak (reflectance dip $R_1(\omega)$) at the 0.5 level amounts to $\Delta_1 = |\text{Im}\,\omega_p| = 4.2 \cdot 10^{12}\,\text{i}\,\text{s}^{-1}$ (~1 nm-wide wavelength interval).

Figure 3 shows the calculated reflectance and transmittance spectra of the composite structures consisting of $N$ W-structures for several $N$ values (solid lines). For comparison, dotted lines show the spectra of the initial W-structure. Figures 3(a) and 3(b) show the spectra of the composite structures consisting of two and four W-structures, respectively. The thicknesses of the spacer layers between the W-structures are determined by Eqs. (10) and (16). In this case, according to the results obtained in Section 2, the complex transmission coefficients of the composite W-structures coincide with the transfer functions of the Butterworth filter of the second and third orders. The transmittance spectra $T_N(\omega) = |t_N(\omega)|^2$ are described by Eqs. (14) (at $N = 2$) and (19) (at $N = 4$). The presented simulation results [Figs. 3(a) and 3(b)] are in full accordance with the theoretical description derived in Section 2. Indeed, the squared moduli of the transfer function of the Butterworth filters of Eqs. (14) and (19), which are shown with dashed lines in Fig. 2, almost completely coincide with the calculated transmission spectra. Note that the widths of the resonant features (transmittance peaks and reflectance dips) of the composite structures are less than the resonance width of the initial W-structure by $\sqrt{2}$ [Fig. 3(a)] or by 2 [Fig. 3(b)] times.

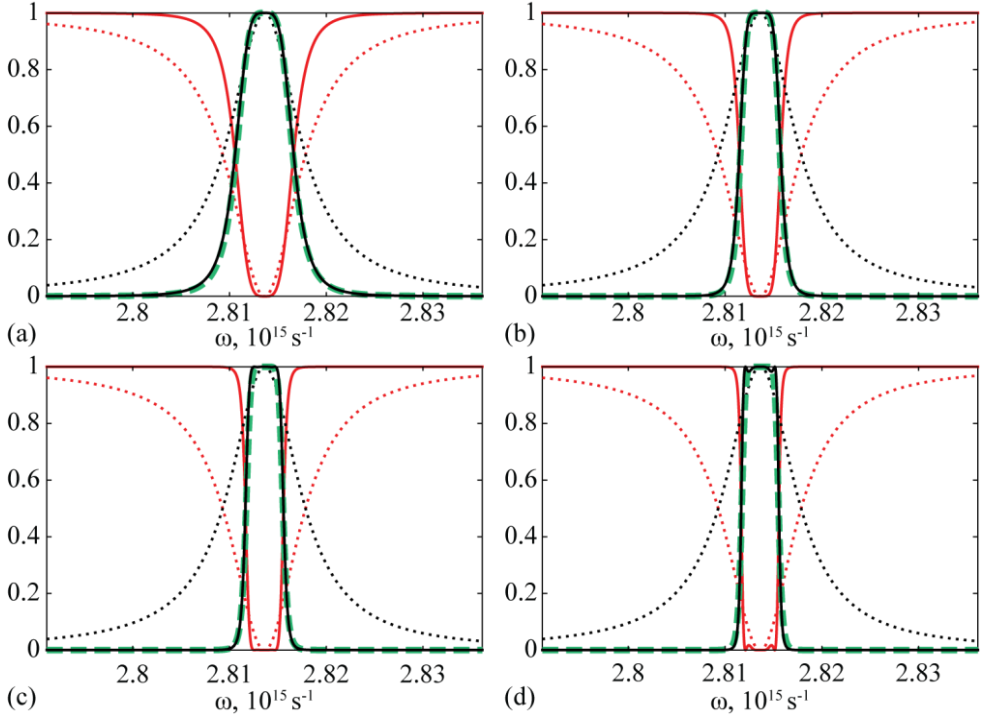

Fig. 3. Reflectance (red) and transmittance (black) of the composite W-structures with $h_{cl} = 700$ nm (solid lines) at $N = 2$ (a), $N = 4$ (b), $N = 6$ (c), and $N = 8$ (d). Dotted lines show the spectra of the initial W-structures. Dashed green lines show the squared moduli of the transfer functions of the corresponding Butterworth filters.

Figures 3(c) and 3(d) show the spectra of the composite structures consisting of six and eight W-structures and described by Eqs. (20) and (24), respectively. The transmittance spectra of these structures are also in good agreement with the theoretical spectra of Eq. (13) at $M = 4$, $M = 5$ and $\Delta = 0.42\Delta_1$. The latter are shown in Figs. 3(c) and 3(d) with dashed lines. However, the difference between the calculated and theoretical spectra is slightly larger in Figs. 3(c) and 3(d) than in Figs. 3(a) and 3(b). In particular, the transmittance peak in Fig. 3(d) is not entirely flat-top and has slight variations with the amplitude of about 0.015. This is due to the fact that the composite structures of Eqs. (20) and (24) only approximately correspond to the Butterworth filters of the fourth and fifth orders. Despite the differences of the simulated spectra in Figs. 3(c) and 3(d) from the spectra of the corresponding Butterworth filters, the composite structures with $N = 6$ [Fig. 3(c)] and $N = 8$ [Fig. 3(d)] still enable high-quality almost flat-top filtering profiles with steep slopes and low sidebands.

The width of the transmittance peak of the composite structure is determined by the width of the resonance of the initial W-structure. In the example considered above ( $h_{cl} = h_{cl,1} = 700$ nm ), the half-width of the transmittance peak of the initial W-structure amounts to $\Delta_1 = \left|\operatorname{Im}\omega_p\right| = 4.2 \cdot 10^{12}\mathrm{i}\ \mathrm{s}^{-1}$, which corresponds to an approximately 1 nm-wide wavelength range. For the considered composite structures, the half-width of the peak changes from $\Delta_1/\sqrt{2}$ (0.71 nm-wide wavelength range) at $N = 2$ [Fig. 3(a)] to $0.46\Delta_1$ (0.46 nm-wide wavelength range) at $N = 8$ [Fig. 3(d)]. For the initial W-structure, the resonance width $\Delta_1 = \left|\operatorname{Im}\omega_p\right|$ is an exponentially decaying function of the quantity $\gamma = h_{cl}/h_{core}$ [13]. Therefore, by changing the cladding thickness $h_{cl}$, one can obtain an arbitrarily narrow transmittance peak (Fig. 2). Consequently, using a composite structure consisting of several identical RDSs, it is possible to achieve a flat-top peak with a required small width. As an example, Fig. 4 shows the transmittance spectra of the composite structures with $N = 4$ and $N = 6$ consisting of W-structures with different thicknesses of the cladding layers $h_{cl,1} = 700\,\text{nm}$, $h_{cl,2} = 800\,\text{nm}$ and $h_{cl,3} = 1000\,\text{nm}$. Let us note that the transmittance spectra at $h_{cl,1} = 700\,\text{nm}$ shown in Fig. 4 coincide with the transmittance spectra in Figs. 3(b) and 3(c), where a wider frequency range is shown. From Fig. 4, it is evident that a decrease in the resonance width caused by an increase in the $h_{cl}$ value is in accordance with Fig. 2. In particular, the width of the resonance of the W-structure in Fig. 2 at $h_{cl} = h_{cl,3} = 1000$ nm amounts to $\Delta_1 = 2.15 \cdot 10^{12}\,\text{i}\,\text{s}^{-1}$ (0.051 nm), which is approximately 20 times less than at $h_{cl} = h_{cl,1} = 700\,\text{nm}$. Similarly, the half-widths of the transmittance peaks of the composite structures at $h_{cl} = h_{cl,3} = 1000$ nm are also approximately 20 times less than at $h_{cl} = h_{cl,1} = 700\,\text{nm}$ and equal $1.08 \cdot 10^{11}\,\text{i}\,\text{s}^{-1}$ (0.026 nm-wide wavelength range) at $N = 4$ and $9.96 \cdot 10^{10}\,\text{i}\,\text{s}^{-1}$ (0.024 nm-wide wavelength range) at $N = 6$.

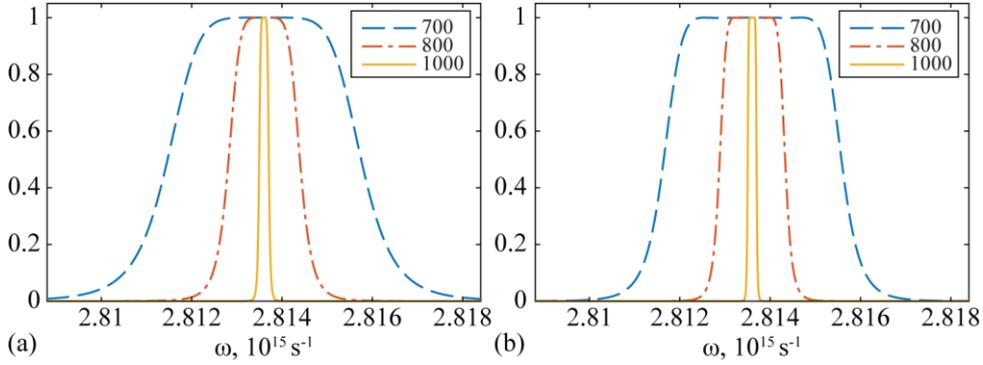

Fig. 4. Transmittance spectra of the composite W-structures with $N = 4$ (a) and $N = 6$ (b) calculated for different thicknesses of the cladding layers: $h_{cl,1} = 700\,\text{nm}$ (dashed blue line), $h_{cl,2} = 800\,\text{nm}$ (dash-dotted red line), and $h_{cl,3} = 1000\,\text{nm}$ (solid yellow line).

## 4. Conclusion

In the present work, we theoretically investigated resonant properties of composite structures consisting of several identical resonant diffractive structures (RDSs) with a Lorentzian transmittance spectrum. Using the scattering matrix formalism, we have analytically demonstrated that the complex transmission coefficients of the composite structures consisting of two and four RDSs coincide with the transfer functions of the Butterworth filters of the second and third orders, respectively. In addition, it has been shown that the composite structures consisting of three and four second-order Butterworth filters approximate fourth- and fifth-order Butterworth filters. The obtained theoretical results are fully confirmed by the results of rigorous numerical simulations of composite W-structures. The simulation results demonstrate the formation of flat-top transmission spectra, the shape of which is in good agreement with the presented theoretical description. Moreover, it is possible to control the width of the flat-top transmission peak by changing the thickness of the cladding layers of the initial W-structure. The presented examples demonstrate the possibility of obtaining flat-top transmission peaks with a significantly subnanometer size.

In addition to the applications in spectral filtering, the proposed composite structures can be used for performing the transformations of optical signals. In particular, the complex reflection coefficients of the composite structures consisting of two and four RDSs have real-valued zeros of the order two and three, respectively. This makes it possible to utilize such structures for the optical computation of second- and third-order derivatives of the envelope of the incident optical pulse in reflection regime [25].

**Funding**

Russian Science Foundation (19-19-00514); Russian Federation Ministry of Science and Higher Education (State contract with the "Crystallography and Photonics" Research Center of the RAS under agreement 007-GZ/Ch3363/26).

**Acknowledgments**

The theoretical investigation of the resonant properties of the composite structures (Section 2) was supported by Russian Science Foundation; the studies regarding the numerical investigation of composite W-structures (Section 3) were supported by the Russian Federation Ministry of Science and Higher Education.